\documentclass[a4paper,aps,pra,preprintnumbers]{revtex4-2}
\usepackage[utf8]{inputenc}
\usepackage[english]{babel}
\usepackage[T1]{fontenc}
\usepackage{amssymb,amsfonts,amsmath,mathtext,enumerate,float,dsfont}
\usepackage{graphics,graphicx,epsfig,epstopdf}
\usepackage{caption}
\usepackage{cmap}
\usepackage{multirow}
\usepackage{indentfirst}
\usepackage[usenames]{color}
\usepackage{amsthm}
\usepackage{xcolor}
\usepackage{ulem}

\renewcommand{\Im}{\mathop{\mathrm{Im}}\nolimits}
\renewcommand{\Re}{\mathop{\mathrm{Re}}\nolimits}

\begin{document}
\title{Growth of Schrödinger cats in particle-number measurement schemes}
\author{S. B. Korolev$^{1}$}
\author{A. A. Silin$^{1}$} 
\author{A. A. Poshevkina$^{2,3}$}
\author{T. Yu. Golubeva$^{1}$}
\affiliation{$^1$ St. Petersburg State University, Universitetskaya nab. 7/9, St. Petersburg, 199034, Russia}
\affiliation{$^2$ Moscow Institute of Physics and Technology,  Institutsky lane 9, Dolgoprudny, 141700, Russia}
\affiliation{$^3$ Russian Quantum Center, Skolkovo IC, Bolshoy Bulvar 30, bld. 1, Moscow, 121205, Russia}

\begin{abstract}
In this work, we investigate the generation of squeezed Schrödinger cat states in schemes based on photon-number-resolving measurements on multimode Gaussian states. We derive analytical expressions for the states generated in two- and three-mode schemes, as well as formulas for their fidelity with squeezed Schrödinger cat states. We analyze how the amplitude of the generated states scales with the number of detected particles. Furthermore, we derive an upper bound on the achievable generation fidelity and identify the conditions under which multimode schemes can enhance the quality of the generated states.
\end{abstract}
\maketitle

\section{Introduction}
Non-Gaussian quantum states play an important role in modern quantum information protocols. In particular, they find applications in problems of quantum metrology \cite{Zhang2019,Hou2019,Munoz2021}, teleportation \cite{Opatrny2000,Zinatullin2023,Zinatullin2021}, and cryptography \cite{Lee2019, Guo2019}. Non-Gaussian states are also necessary for the implementation of universal quantum computation  \cite{Braunstein_2005,Lloyd_1999}. The nonclassical statistics and asymmetries of non-Gaussian states allow them to be used in quantum error-correction codes \cite{Hastrup2023,Korolev2024_eror_cor, Binomial_state,Bashmakova2025,Korolev2026}.

Among non-Gaussian states, Schrödinger cat states \cite{YurkeStoler1986} attract special attention, since they have a wide range of applications. They are used for fault-tolerant quantum computation \cite{Ralph_2003,Mirrahimi_2014}, as well as for protecting information against particle-loss errors \cite{Cochrane1999,Bergmann2016,Mirrahimi_2014}. Squeezed Schrödinger cat states (SSCSs) are of additional interest. Such states combine the nonclassical interference properties of a superposition with the suppression of quantum fluctuations in one of the quadratures. This opens up additional possibilities for using these states in various quantum information protocols \cite{Marek2008,schlegel2022quantum,hillmann_2023}.

The experimental generation of non-Gaussian states in the optical domain remains a challenging task. Deterministic generation methods based on nonlinear interactions are difficult to implement due to weak optical nonlinearities \cite{Munro2005}. For this reason, quantum-state engineering based on conditional measurements (heralded generation) \cite{Xue2015,Eaton2019,Kuts2022,Chen2024,Korolev2024fock} has become one of the most efficient approaches for generating non-Gaussian states in optics. In particular, photon-number measurements on multimode Gaussian states constitute a flexible and experimentally accessible method for generating a wide class of non-Gaussian states \cite{Su2019,Gagatsos2021,Podoshvedov2023}.

Heralded schemes based on photon-number measurements on two-mode Gaussian entangled states have been actively studied for generating various non-Gaussian states of light. Moreover, in \cite{Takase2021} it was demonstrated that such schemes are also applicable for generating approximate Schrödinger cat states. However, many important open questions remain. In particular, what are the best parameters that the SSCSs generated in two-mode schemes can possess; how does increasing the number of modes in the schemes affect the parameters of the generated states; and whether there exists a fundamental limit on the parameters of the generated SSCS in the most general multimode schemes based on particle-number measurements.

In this work, we systematically investigate the heralded generation of squeezed Schrödinger cat states using two- and three-mode Gaussian states. We show that efficient generation of SSCSs in two-mode schemes requires an appropriate choice of scheme parameters. In addition, we demonstrate that increasing the number of modes in the schemes improves the fidelity of the generated SSCSs, but reduces the probability of obtaining them. We derive an upper bound on the fidelity of SSCSs generated in arbitrary multimode schemes based on particle-number measurements in the modes of a multimode Gaussian state.

The paper is organized as follows. In Section II we show how to decompose the SSCS in a form convenient for analysis and derive a fundamental upper bound on the fidelity of generating such states in multimode schemes based on measurements on Gaussian states. In Sections III and IV, we investigate the characteristics of SSCSs generated in two-mode and three-mode schemes, respectively. In Section IV, we also compare the two schemes in terms of the quality of the generated SSCSs.

\section{Generation of squeezed Schrödinger cat states in schemes based on particle-number measurements of multimode Gaussian states}
\subsection{Squeezed Schrödinger cat state}
Before estimating the fidelity of SSCS generation in schemes based on particle-number measurements of multimode Gaussian states, we first define these states. In this work, we consider SSCSs specified by the following state vector:
\begin{align} \label{cat_def}
    |\text{Scat}_{\pm} \left(\alpha,R\right)\rangle=\frac{1}{\sqrt{N_{\text{Scat}_{\pm}}}}\left(|\alpha,R\rangle\pm|-\alpha,R\rangle\right), 
\end{align}
where $N_{\text{Scat}_{\pm}}$ is the normalization factor, $|\alpha,R\rangle=\hat{\mathcal{S}}\left(R\right)\hat{\mathcal{D}}\left(\alpha\right)|0\rangle$  is a squeezed coherent state, $|0\rangle$ is the vacuum state, $\hat{\mathcal{D}}\left(\alpha\right)=\exp \left[\alpha \hat{a}^{\dag}-\alpha^* \hat{a}\right]$ is the displacement operator, and $\hat{\mathcal{S}}\left(R\right)=\exp\left[\frac{R}{2}\left(\left(\hat{a}^\dag\right)^2-\hat{a}^2\right)\right]$ is the squeezing operator. The signs $+$ and $-$ in expression (\ref{cat_def}) correspond to the even and odd SSCS, respectively. In this work, we consider SSCSs with real parameters $\alpha, R \in \mathbb{R}$.

For further analysis, it is convenient to rewrite the SSCS as an expansion over squeezed Fock states. Such expansions look as follows:
\begin{align}
    &|\text{Scat}_{+}(\alpha,R)\rangle=\hat{\mathcal{S}}\left(R\right) \left(\sum _{m=0}^{\infty}\frac{\alpha^{2 m}}{\sqrt{ (2m)!\cosh \left(\alpha^2\right)}}|2m\rangle\right), \label{cat_pl}\\
    &|\text{Scat}_{-}(\alpha,R)\rangle= \hat{\mathcal{S}}\left(R\right) \left(\sum_{m=0}^{\infty}\frac{\alpha^{2m+1} }{\sqrt{(2 m+1)!} \sqrt{\sinh \left(\alpha^2\right)}}|2m+1\rangle\right), \label{cat_min}
\end{align}
From the expansion it is clear that SSCSs are decomposed as an infinite superposition over squeezed Fock states with squeezing parameter $R$.

In addition, for further analysis, it is convenient to expand the SSCS under study in terms of squeezed Fock states with a different squeezing parameter $r$. Since squeezed Fock states form a complete orthonormal set, any state can be expanded in this basis.  The expansion of an SSCS with squeezing parameter $R$ in the basis of squeezed Fock states with squeezing parameter $r$ is given by:
\begin{align} 
  &|\text{Scat}_{+}(\alpha,R)\rangle =\hat{\mathcal{S}}(r)\left(\sum_{m=0}^{\infty} \frac{ \left(\frac{\tanh (R-r)}{2}\right)^m e^{\frac{1}{2} \alpha^2 \tanh (R-r)} H_{2
   m}\left(\frac{\alpha}{\sqrt{\sinh (2 (R-r))}}\right)}{\sqrt{(2 m)! \cosh \left(\alpha^2\right) \cosh (R-r)}}  |2m\rangle \right), \label{SSCS_dec_SFS_1}\\
  &|\text{Scat}_{-}(\alpha,R)\rangle = \hat{\mathcal{S}}(r)\left(\sum_{m=0}^{\infty}\frac{ \left(\frac{\tanh (R-r)}{2}\right)^{m+\frac{1}{2}} e^{\frac{1}{2}
   \alpha^2 \tanh (R-r)} H_{2
   m+1}\left(\frac{\alpha}{\sqrt{\sinh (2 (R-r))}}\right)}{\sqrt{(2
   m+1)! \sinh \left(\alpha^2\right) \cosh (R-r)}}|2m+1\rangle\right). \label{SSCS_dec_SFS_2}
\end{align}
where $H_n$ stands for the Hermite polynomials defined such that the three  lowest of them read
\begin{equation}
\label{Her}
H_0(x) = 1, \qquad  H_1(x) = 2x,\qquad \mathrm{and}  \qquad H_2(x) = 4x^2-2.
\end{equation} 
From the expansions obtained, it is clear that for real $\alpha$ and $R$ the expansion coefficients of the SSCS over the squeezed Fock states are also real.

\subsection{Maximal fidelity of SSCSs in schemes based on particle-number measurements on a multimode Gaussian state}
In the previous subsection, we showed that SSCSs can be expanded as an infinite superposition of squeezed Fock states of definite parity. It follows that any state close to an SSCS should have a similar expansion.

In this work, we use states obtained in schemes based on particle-number measurements performed on multimode Gaussian states to approximate SSCSs. As was shown in \cite{Su2019}, such schemes generate states of the form:
\begin{align} \label{opt_superpos}
    |\Phi _n(r) \rangle= \hat{\mathcal{S}}(r) \left( \sum \limits_{k=1}^{\lfloor n/2\rfloor} c_{k}|n-2k\rangle \right).
\end{align}
where $n$ is the total number of particles detected. From this expansion it follows that, depending on the parity of the detected number of particles $n$, the state expands into either even or odd squeezed Fock states. In addition, the output state is a finite superposition of squeezed Fock states. Such states can approximate any SSCS for appropriate choices of the expansion coefficients  $c_k$ and the detected particle number $n$. However, since the expansion is finite-dimensional, this approximation is limited. 

Let us estimate this limitation. To this end, we formulate a general bound on the maximal fidelity between states with finite and infinite expansions in a given orthonormal basis. Let $|\Psi\rangle$ be an arbitrary state whose expansion in a complete orthonormal basis $|\phi_k \rangle$ has the following form:
\begin{align}
    |\Psi\rangle=\sum_{k=0}^{\infty} a_k |\phi_k\rangle,
\end{align}
where  $a_k=\langle \phi_k|\Psi\rangle$ are the expansion coefficients satisfying the condition $\sum \limits_{k=0}^{\infty} |a_k|^2=1$. The controllable state $|\Psi_n\rangle$, which we want to bring close to the state $|\Psi\rangle$, is a finite superposition of the basis states $|\phi_k \rangle$:
\begin{align}
    |\Psi_n\rangle=\sum_{k=0}^{n} b_k |\phi_k\rangle,
\end{align}
where the expansion coefficients satisfy the normalization condition $\sum \limits_{k=0}^{n} |b_k|^2=1$. 

The fidelity of two such states is given by the following expression:
\begin{align} \label{fid_theor}
    F\left(|\Psi_n\rangle,|\Psi\rangle\right)=\left|\langle \Psi|\Psi_n \rangle\right|^2=\left|a_1^*b_1+a_2^*b_2+\dots +a_n^*b_n\right|^2.
\end{align}
Here we have used the fact that one of the states has a finite expansion in the chosen orthonormal basis. The fidelity approaches unity as the two states become increasingly similar. To bring the controllable state close to the desired one, we should choose the coefficients $b_k$ such that the fidelity is maximal. To determine the maximal fidelity, we derive an upper bound on Eq. (\ref{fid_theor}) using the Cauchy–Bunyakovsky–Schwarz inequality. Since both states are normalized, we obtain the following inequality:
\begin{align}
F\left(|\Psi_n\rangle,|\Psi\rangle\right)=\left|a_1^*b_1+a_2^*b_2+\dots +a_n^*b_n\right|^2 \leqslant \sum _{k=0}^{n}|a_k|^2 \sum _{k=0}^{n}|b_k|^2=\sum _{k=0}^{n}|a_k|^2.
\end{align}
Thus, we have obtained an upper bound for the value of the fidelity: $F\leqslant \sum \limits _{k=0}^{n}|a_k|^2$. The equality in this expression is achieved for the following choice of parameters of the state $|\Psi_n\rangle$:
\begin{align}
    b_k=\frac{a_k}{\sqrt{\sum \limits _{k=0}^n}|a_k|^2}.
\end{align}
In this case, the fidelity is maximized and is equal to $F_{\text{max}}\left(|\Psi_n\rangle,|\Psi\rangle\right)= \sum \limits _{k=0}^{n}|a_k|^2$. 
It follows that the state closest to $|\Psi \rangle$ is 
\begin{align}
    |\Psi_n\rangle= \sum_{k=0}^n\frac{\langle \phi_k|\Psi\rangle}{\sqrt{\sum \limits _{k=0}^n|\langle \phi_k|\Psi\rangle|^2}} |\phi_k\rangle, 
    \end{align}
and the corresponding maximum fidelity is
\begin{align}
    F_{\text{max}} \left(|\Psi_n\rangle,|\Psi\rangle\right)=\max_{|\Psi_n\rangle} |\langle\Psi_n|\Psi \rangle|^2= \sum \limits _{k=0}^{n}|\langle \phi_k|\Psi\rangle|^2.
\end{align}

In the case we are studying, the controllable states are finite superpositions of squeezed Fock states with definite parity, so that  $|\phi _k\rangle =\hat{\mathcal{S}}(r)|k\rangle$. Using the expansions of the SSCSs given in Eqs. (\ref{SSCS_dec_SFS_1}) and (\ref{SSCS_dec_SFS_2}), we obtain the maximum fidelity between an SSCS and a finite superposition of squeezed Fock states:
\begin{multline} \label{opt_fid_0}
F_{\text{max}} \left(|\Phi _n(r) \rangle,|\text{Scat}_{\pm}(\alpha,R)\rangle\right)= \frac{2 e^{\alpha^2 \tanh(R - r)}}{\left(e^{\alpha^2}+(-1)^n e^{-\alpha^2}\right) \cosh(R - r)}\\
   \times \sum_{k=0}^{\lfloor n/2 \rfloor} \frac{\left( \frac{\tanh(R - r)}{2} \right)^{n - 2k} H_{n-2k}^2 \left( \frac{\alpha}{\sqrt{\sinh(2R - 2r)}} \right)}{(n - 2k)!},
\end{multline}
In evaluating the fidelity of the states, we obtained a unified expression that covers both cases (even and odd SSCS). It depends on the SSCS parameters (amplitude $\alpha$ and squeezing $R$) and on the parameters of the controllable state (squeezing $r$ and number of terms in the expansion $n$).

\begin{figure}[H]
    \centering
    \includegraphics[width=0.5\linewidth]{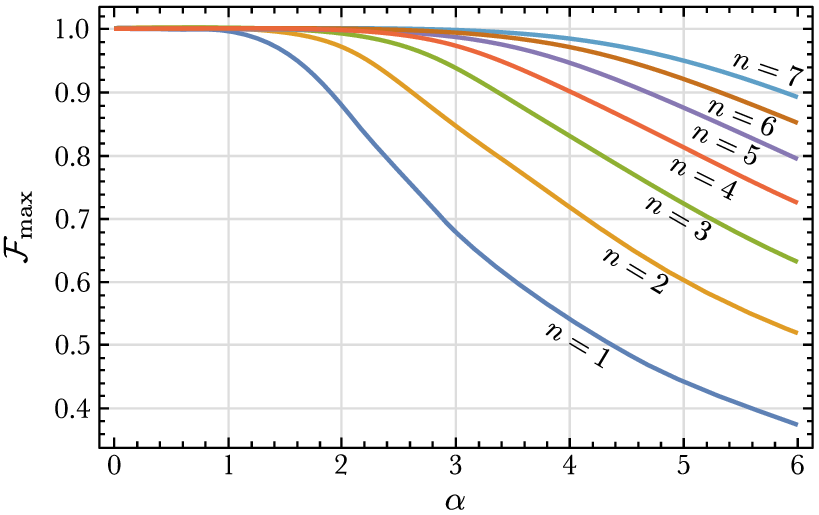}
    \caption{Maximal multimode fidelity of SSCS generation as a function of the amplitude $\alpha$. Different colors denote different cases of the detected number of particles $n$.}
    \label{fig:optimal}
\end{figure}

It is important to note that the fidelity depends only on the difference between the squeezing parameters $R-r$. We therefore introduce a new parameter, $y=\tanh \left(R-r\right)$, which allows Eq. (\ref{opt_fid_0}) to be written in a simpler form. The parameter $y$ characterizes the squeezing of the controllable state $|\Phi (r)\rangle$ relative to the squeezing of the generated SSCS. This is a controllable parameter over which we can maximize the fidelity of generating an SSCS with amplitude $\alpha$ and any squeezing $R$. Consequently, the maximum fidelity depends only on the amplitude $\alpha$  and is given by
\begin{align} \label{opt_fid}
 \mathcal{F}_{\text{max}}\equiv\max_y \left[\frac{2 e^{\alpha^2 y} \sqrt{1 - y^2}}{e^{\alpha^2}+(-1)^n e^{-\alpha^2}} \sum_{k=0}^{\lfloor n/2 \rfloor} \frac{\left( \frac{y}{2} \right)^{n - 2k} H_{n-2k}^2 \left( \frac{\alpha}{\sqrt{2}} \frac{\sqrt{1 - y^2}}{\sqrt{y}} \right) }{(n - 2k)!}\right].
\end{align}
We refer to this case as the optimal case and to the corresponding fidelity as the maximal multimode fidelity. Fig. \ref{fig:optimal} presents the maximal multimode fidelity as a function of the SSCS amplitude $\alpha$.
\noindent As shown in Fig. \ref{fig:optimal}, high-fidelity SSCS generation requires states of the form (\ref{opt_superpos}) with a large number of superposition terms, obtained by detecting a large total number of particles $n$. However, the probability of generating such states in realistic schemes is generally low \cite{Fiurasek2026}.

The presented plot also helps us to understand the constraints on the SSCSs generated in potential schemes. For example, from Fig. \ref{fig:optimal} one can conclude that in any scheme, upon detecting a total of two particles in the modes of a multimode Gaussian state, it is impossible to generate an SSCS with amplitude $\alpha=3$ and fidelity above $0.9$.

The fidelity estimates presented above provide an upper bound achievable in schemes based on particle-number measurements in multimode Gaussian states. In practical implementations, this bound is not always attainable due to the limited number of controllable parameters. However, as the number of modes increases, so does the number of controllable parameters, and the achievable fidelity approaches this upper bound. In the following, we consider two generation schemes based on two- and three-mode Gaussian states and estimate how closely the resulting states approximate an SSCS.

\section{Generation of SSCSs in schemes with a two-mode Gaussian state}
\subsection{States in a two-mode scheme with particle-number measurements}
\begin{figure}[H]
    \centering
    \includegraphics[scale=0.8]{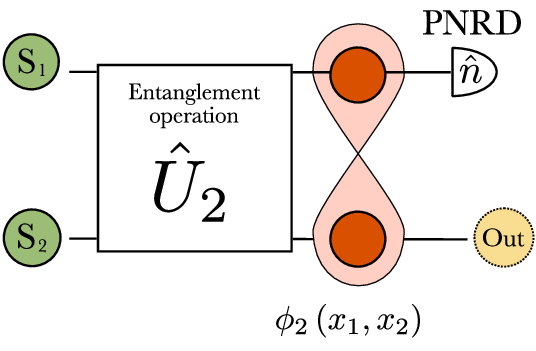}
    \caption{Two-mode scheme for generating SSCS. In the figure, $\text{S}_1$ and $\text{S}_2$ are two squeezed vacuum states, $\hat{U}_2$ is some two-mode entanglement Gaussian operation, PNRD is a photon-number-resolving detector, and Out is the output state.} 
    \label{fig:scheme_2}
\end{figure}

We now turn to the study of states generated via particle-number measurements on one mode of a two-mode Gaussian state. We consider the scheme shown in Fig. \ref{fig:scheme_2}. A similar approach to generating SSCSs was previously studied in \cite{Takase2021}. Unlike that work, we treat the generated states in their most general form without introducing any approximations. As shown below, this leads to the generation of SSCSs with higher fidelity.
\noindent As shown in the figure, two arbitrary squeezed vacuum states are entangled via a two-mode Gaussian transformation $\hat{U}_2$. We do not specify the form of this transformation; it may correspond to a beam splitter, a CZ transformation, or some other transformation. Following this transformation, the two initially independent states become a two-mode Gaussian state. In the coordinate representation, this state can be written in the general form
\begin{align}
\label{TMGS}
   \phi_2\left(x_1,x_2 \right)=\frac{\left(\mathrm{Re}[ a] \mathrm{Re} [c]-\left(\mathrm{Re }[b] \right)^2\right)^{1/4}}{\sqrt{\pi}}\exp \left[{-\frac{1}{2} \left(a x_1^2+2 b x_1 x_2+c x_2^2\right)}\right],
\end{align}
where $a,b,c  \in \mathbb{C}$ are variables that are certain functions of the three parameters of the generation scheme: the squeezing of the input oscillators and the entanglement operation \cite{Korolev2024fock,Takase2021}. For the function (\ref{TMGS}) to be the wave function of a real physical system, the parameters must satisfy the following conditions:
\begin{align}
 \label{param_TMSS}
 \mathrm{ Re}[a] >0, \qquad   \mathrm{ Re}[c] >0, \qquad  \mathrm{Re} [a] \mathrm{Re}[ c]-\left(\mathrm{Re }[b] \right)^2 >0.
\end{align}

After measuring one mode of this state using a photon-number-resolving detector (PNRD), a non-Gaussian state is obtained in the remaining mode \cite{Korolev2024estimation}, described by the following state vector:
\begin{align} \label{out_Rz}
  |\Psi_n^{\text{out},2} (r,z)\rangle  =\hat{\mathcal{S}}\left(|r| e^{i\arg [r]}\right)\left(\sum \limits_{k=0}^{\lfloor\frac{n}{2} \rfloor}  c_{k}(z) |n-2k\rangle\right),
\end{align}
where the expansion coefficients are given by the following expressions:
\begin{align} \label{coef_2mode}
    c_{k}\left(z\right)= \frac{z ^{k}}{ 2^{k} k! \sqrt{\,
   _2F_1\left(\frac{1-n}{2},-\frac{n}{2};1;\left| z\right| ^2\right)}}\sqrt{\frac{n!}{(n-2k)!}}.
\end{align}
Here $ _2F_1(x,y;s;t)$ is the hypergeometric function. In the presented expansion, two parameters are used, $r=|r|e^{i \arg [r]}$ and $z$, which are related to the parameters of the two-mode Gaussian state as follows:
\begin{align} \label{param_N2}
     z=1-\frac{\left(a^2-1\right)}{b^2}\Re\left[c-\frac{b^2}{a+1}\right],\quad \cosh (|r|)=\frac{\left| 1+c-\frac{b^2}{a+1}\right| }{2 \sqrt{\Re\left[c-\frac{b^2}{a+1}\right]}},\quad \tan (\arg \left[r\right])=\frac{2 \Im\left[c-\frac{b^2}{a+1}\right]}{\left| c-\frac{b^2}{a+1}\right| ^2-1}.
\end{align}
It follows from the above expansion that the generated state is characterized by two parameters. The parameter $r$ determines the squeezing of the state, while $z$ determines the coefficients of its Fock-state expansion. Since both parameters are related to the scheme parameters through Eq. (\ref{param_N2}), varying the squeezing of the input modes and the entanglement operation enables the generation of a broad class of non-Gaussian states \cite{Korolev2024estimation}.

It is also important to note here that the three scheme parameters $a$, $b$, and $c$ map into two parameters of the generated state: $r$ and $z$. This means that one free parameter remains, which does not affect the generated state but affects its generation probability. This question will be investigated in more detail later.

\subsection{Fidelity of SSCSs generated in the scheme}
Using the analytical expansions of the SSCSs in the squeezed Fock-state basis (\ref{SSCS_dec_SFS_1}) and (\ref{SSCS_dec_SFS_2}), we can directly compare them with the states generated in the scheme shown in Fig. \ref{fig:scheme_2}. Because the expansion coefficients of the considered SSCSs are real, the coefficients (\ref{coef_2mode}) must also be real. Accordingly, we restrict the parameters to real values, $z \in \mathbb{R}$ and $r \in \mathbb{R}$.

With the constraints $z \in \mathbb{R}$ and $r \in \mathbb{R}$ imposed, we quantify the similarity between the two states via the fidelity:
\begin{align} \label{fid}
 F\left(|\Psi_n^{\text{out},2}(r,z) \rangle,|\text{Scat}_{\pm}(\alpha,R)\rangle\right) =  \frac{2 e^{\alpha^2 y} \sqrt{1-y^2}  \left(\frac{y+z}{2}\right)^n H_n^2\left(\frac{\alpha}{\sqrt{2}}\frac{\sqrt{1-y^2}}{\sqrt{y+z}}\right)}{n! \left((-1)^n e^{-\alpha^2}+e^{\alpha^2}\right) \,
   _2F_1\left(\frac{1-n}{2},-\frac{n}{2};1;z^2\right)}.
\end{align}
Here it is taken into account that depending on the parity of $n$, we compare our state with an even or odd SSCS. In addition, in the presented expression, as before, we introduce the parameter $y$, which is responsible for the amount of squeezing of the generated state relative to the squeezing of the SSCS. The expression also contains the parameter $z$, responsible for the superposition coefficients of our state. By choosing these two parameters, we can bring the states generated in the scheme shown in Fig. \ref{fig:scheme_2} upon detecting $n$ particles close to an SSCS with amplitude $\alpha$ and any amount of squeezing $R$.
 
\begin{figure}[H]
    \centering
    \includegraphics[width=0.6\linewidth]{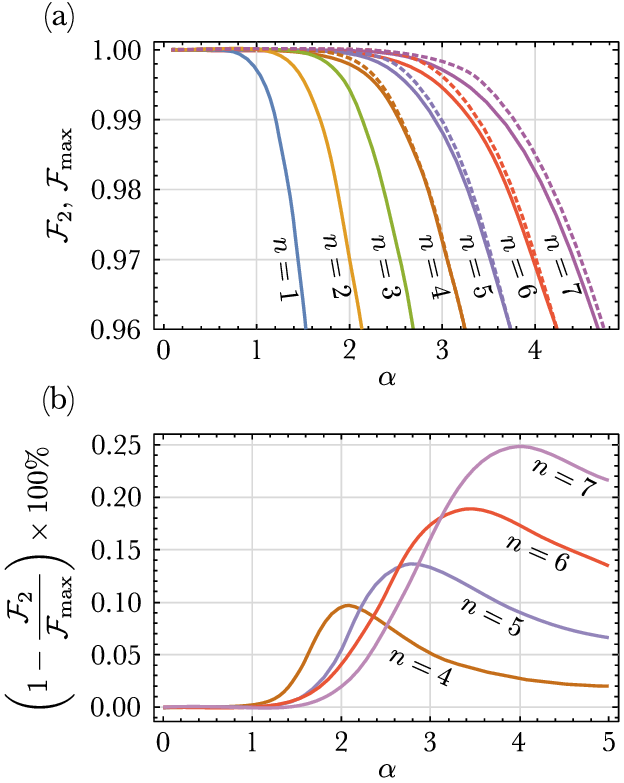}
    \caption{(a) Maximal SSCS generation fidelity as a function of the amplitude $\alpha$ in two cases: in the case of using optimal states of a finite superposition of squeezed Fock states (dashed line) and the states obtained in the scheme based on measurement of a two-mode Gaussian state (solid line). (b) The relative fidelity gain in the two cases under consideration as a function of the SSCS amplitude $\alpha$. Different colors in the plot denote different cases of the detected number of particles n.}
    \label{fig:N2_fid}
\end{figure}

From a physical point of view, the optimization reduces to choosing the scheme parameters that maximize the fidelity of the generated SSCS. Fig. \ref{fig:N2_fid} (a) shows the maximal SSCS fidelity $\mathcal{F}_{2} \equiv \max_{z,y} \left[F\left(|\Psi_n^{\text{out},2}(r,z) \rangle,|\text{Scat}_{\pm}(\alpha,R)\rangle\right) \right]$
as a function of the amplitude $\alpha$  in two cases: for optimal states given by a finite superposition of squeezed Fock states (\ref{opt_superpos}), and for states generated in the scheme based on measurements of a two-mode Gaussian state (\ref{out_Rz}). Fig. \ref{fig:N2_fid} (b)  shows the relative fidelity gain $ \left(1-\frac{\mathcal{F}_2}{\mathcal{F}_{\text{max}}}\right) \times 100 \%$
for the two considerate cases as a function of the SSCS amplitude $\alpha$.
\noindent Fig. \ref{fig:N2_fid} shows that, in the scheme based on particle-number measurements on one mode of a two-mode Gaussian state, the fidelity of the generated SSCS increases with the detected particle number $n$. This is due to the fact that, as $n$ grows, the state (\ref{out_Rz}) acquires more and more superposition terms and approaches more and more closely an SSCS with an infinite number of terms in the expansion.

In addition, from Fig.\ref{fig:N2_fid} (a) it is clear that for $n=1,2,3$, the fidelity of SSCS generation in the two-mode scheme coincides with the maximal fidelity (\ref{opt_fid}). However, starting from $n=4$ the fidelity in the two-mode scheme becomes lower than in the optimal case. Moreover, with increasing $n$, the relative gain of the optimal fidelity over the fidelity in the two-mode scheme also increases. This is due to the fact that in the optimal state the number of controllable parameters is $n-1$, whereas in the state generated in the two-mode scheme there are only two controllable parameters, $z$ and $y$. By controlling these parameters, one can generate an arbitrary superposition state of two squeezed Fock states. Thus, to improve generation fidelity, we should detect a large number of particles $n$, and also increase the number of controllable parameters (the number of modes) in schemes based on particle-number measurements.

It is also important to note that for small detected particle numbers $n$, the difference between the optimal and two-mode cases is small. As shown in Fig. \ref{fig:N2_fid}(b), the maximal relative gain at $n=7$ is only about $0.25 \%$. However, this gain increases rapidly with $n$, and becomes significant for larger values of $n$.

\subsection{Growth of the amplitude of SSCSs generated in the two-mode scheme}
Quantum information protocols often require SSCSs with large amplitudes  $\alpha$  \cite{Ralph_2003,hastrup2022all}. It is therefore important to determine how many particles must be detected in the scheme under consideration to generate states close to an SSCS with a given amplitude. To address this question, we estimate how the achievable SSCS amplitude $\alpha$  scales with the number of detected particles in the scheme shown in Fig. \ref{fig:scheme_2}.

To characterize this scaling, we fix a target value of the maximal fidelity $\mathcal{F}_2$ and determine how the corresponding SSCS amplitude $\alpha$ increases with the detected particle number $n$. Note that the optimization of the scheme parameters must be performed separately for each value of $n$.

\begin{figure}[H]
    \centering
    \includegraphics[scale=0.7]{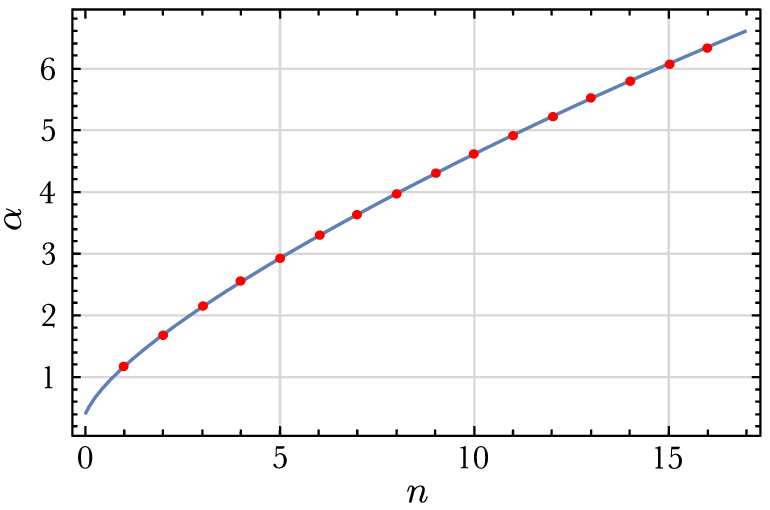}
    \caption{Amplitude $\alpha$ of the generated approximate SSCS with fidelity $0.99$ in the two-mode scheme as a function of the detected particle number $n$. The dots show the numerical values of $\alpha$, while the solid line represents a fit to the data.}
    \label{fig:w_n}
\end{figure}

We first consider a target fidelity of $0.99$. Fig. \ref{fig:w_n} shows the dependence of the SSCS amplitude $\alpha$ on the number of particles detected by the PNRD at this fidelity level.
\noindent Fig. \ref{fig:w_n} shows the numerical values of the achievable SSCS amplitude together with a power-law fit. The fitted dependence is $\alpha\approx 0.40+0.78n^{0.73}$. This fit can be used to estimate the minimum number of detected particles needed to generate an SSCS with arbitrary squeezing $R$ and a given amplitude $\alpha$ in the scheme under consideration.  For example, generating an SSCS with $\alpha=5$ and fidelity $0.99$ requires the detection of approximately 11 particles. Detection of a different particle number leads to a different generated state and, consequently, a different fidelity with respect to the required SSCS.

\begin{figure}[H]
    \centering
    \includegraphics[scale=0.7]{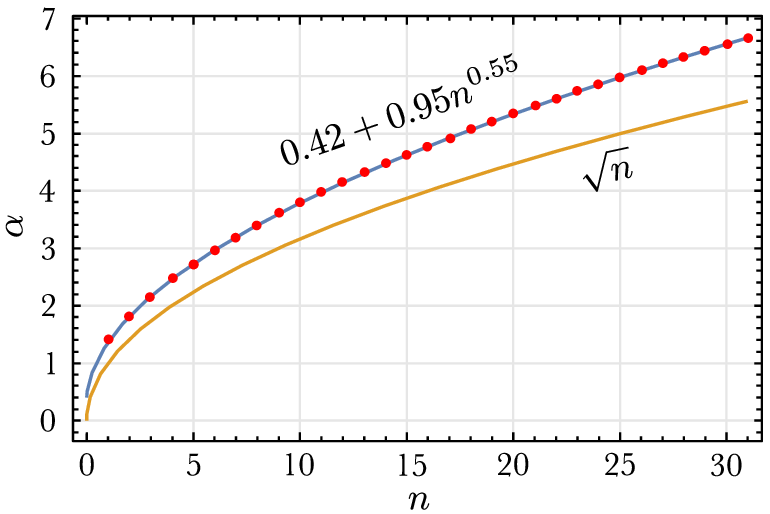}
    \caption{Dependence of the SSCS amplitude $\alpha$  for states generated with fidelity $\mathcal{F}_2=1-\frac{0.03}{n}$ on the detected particle number $n$. The red dots denote the values obtained in the two-mode scheme, while the blue solid line represents its fitting. The orange line shows the dependence $\alpha=\sqrt{n}$ for SSCSs generated in a similar two-mode scheme \cite{Takase2021}.}
    \label{fig:wsqrtn}
\end{figure}

We now compare the scaling of the achievable amplitude in our scheme with that of a similar two-mode scheme without optimizing the scheme parameters for a specific SSCS. In \cite{Takase2021} it was shown that, for increasing detected particle number $n$, states close to an SSCS can be generated with fidelity $F\approx1-\frac{0.03}{n}$ and amplitude $\alpha=\sqrt{n}$. Fig. \ref{fig:wsqrtn} compares the dependence $\alpha=\sqrt{n}$ with that achieved in our scheme at the same fidelity level $\mathcal{F}_2=1-\frac{0.03}{n}$ as a function of the detected particle number.
\noindent The figure shows that the SSCS amplitude at fidelity $\mathcal{F}_2=1-\frac{0.03}{n}$  is well approximated by $\alpha\approx 0.42+0.95 n^{0.55}$. The obtained amplitude exceeds  $\alpha=\sqrt{n}$ for all values of $n$. This demonstrates that optimization of the scheme parameters enables the generation of SSCSs with larger amplitudes for a given detected particle number.

\subsection{Probability of SSCS generation in the scheme based on measurement of a two-mode Gaussian state}
A feature of the generation scheme under consideration is its probabilistic nature. Because the output state (\ref{out_Rz}) is generated upon detection of $n$ particles, the probability of generating a particular state is equal to the probability of observing the corresponding measurement outcome.  This probability is given by the following expression \cite{Korolev2024estimation}:
\begin{align} \label{prob}
    \mathcal{P}_2(n)=\frac{2(a-1)^n}{(a+1)^{n+1}| 1-z| ^{n+\frac{1}{2}}} \sqrt{a(| 1-z| -1)+1} \,
   _2F_1\left(\frac{1-n}{2},-\frac{n}{2};1;| z| ^2\right),
\end{align}
It follows from this expression that the probability depends on the state parameter $z$, as well as on the scheme parameter $a$. The parameter $a$ affects only the generation probability and does not influence the state (\ref{out_Rz}) itself. Consequently, $a$ remains a free parameter that can be chosen to maximize the probability of generating a state with given values of $z$ and $r$.

\begin{figure}[H]
    \centering
    \includegraphics[width=0.65\linewidth]{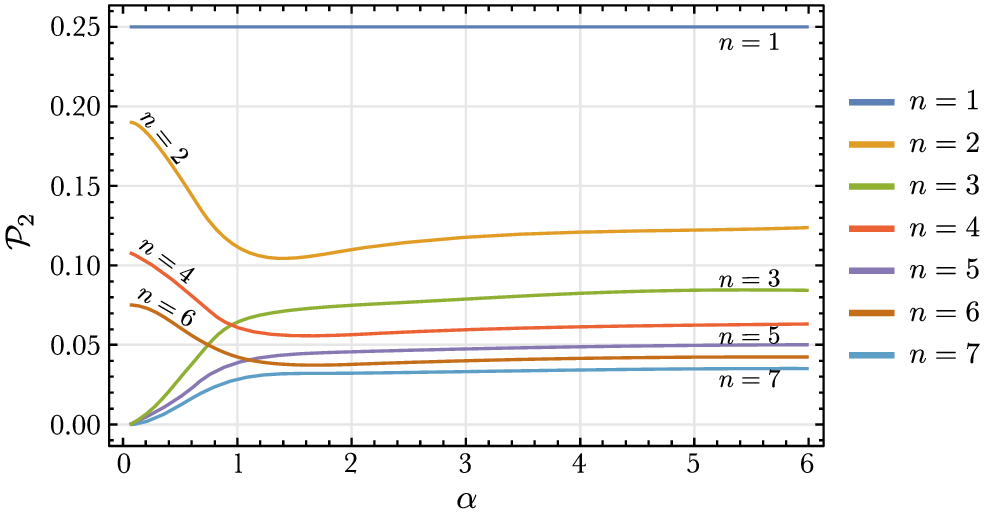}
    \caption{Maximal probability of generating an SSCS with amplitude $\alpha$ in the two-mode scheme at maximal fidelity. Different colors denote the generation probability for different detected particle numbers $n$.}
    \label{fig:prob}
\end{figure}

When generating an SSCS, we choose the parameters $z$, $r$, and $n$ to maximize generation fidelity. Then, knowing these values, we can maximize the probability by choosing the parameter $a$. Fig. \ref{fig:prob} presents plots of the maximal values of the probability of generating an SSCS with amplitude $\alpha$, generated in the two-mode scheme with maximal fidelity.
\noindent Several conclusions can be drawn from the plot. First, for $n=1$, the probability of generating an SSCS is constant and equal to $1/4$. This is because, upon detection of a single particle, the output state of the scheme is always a squeezed first Fock state \cite{Korolev2024fock,Korolev2024estimation}. Such a state is characterized solely by its squeezing and is independent of the parameter $z$. The fidelity can therefore be optimized by adjusting the squeezing (equivalently, the relative squeezing parameter $y$), while the parameter $z$ remains free and can be chosen to maximize the generation probability.

Second, the probability of generating an SSCS with large amplitude decreases with increasing detected particle number $n$. For small amplitudes, the generation probability of even SSCSs, corresponding to even values of $n$, exceeds that of odd SSCSs. However, this probability also decreases as $n$ increases.

These results highlight a trade-off between fidelity and generation probability. On the one hand, high-fidelity generation of large-amplitude SSCSs requires the detection of a large number of particles. On the other hand, such detection events occur with low probability. Therefore, to generate an SSCS in real schemes, one should use the following strategy: specify the target SSCS amplitude $\alpha$, set an acceptable level of fidelity, and determine the smallest number of particles that needs to be detected to generate this state. By following this strategy, one maximizes the probability of generation while maintaining the desired state quality.

\section{Generation of SSCSs in schemes with a three-mode Gaussian state}

\subsection{States in a three-mode scheme with particle-number measurements}
\begin{figure}[H]
    \centering
    \includegraphics[width=0.45\linewidth]{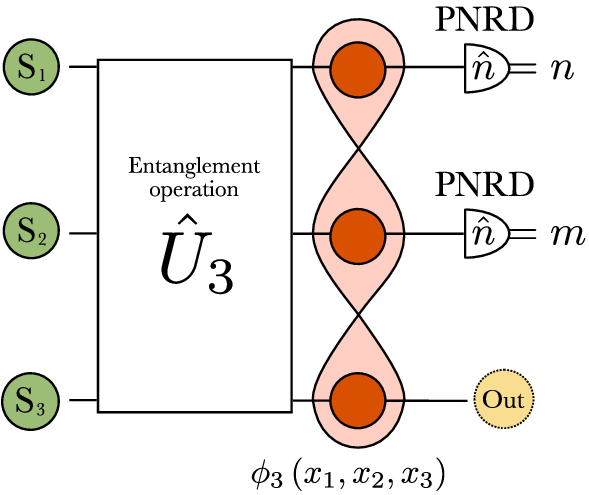}
    \caption{Three-mode scheme for generating SSCSs. In the figure, $\text{S}_1$, $\text{S}_2$  and $\text{S}_3$ are three squeezed vacuum states, $\hat{U}_3$ is some three-mode entangling Gaussian operation, PNRD is a photon-number-resolving detector, and Out is the output state.}
    \label{fig:scheme_3mode}
\end{figure}

In the previous section, we showed that a two-mode particle-number-measurement scheme can generate states close to an SSCS with arbitrary amplitude $\alpha$ and any pre-specified fidelity. However, generating states with large amplitudes requires the detection of a large number of particles $n$. As shown above, for $n\geqslant4$ the fidelity obtained in the two-mode scheme falls below the maximal multimode fidelity $\mathcal{F}_{\text{max}}$ because of the limited number of controllable parameters. Moreover, this discrepancy increases as $n$ increases. Since detecting a large number of particles is low-probability and experimentally challenging, it is desirable for the state generation to be as efficient as possible and to approach the optimal case. This can be achieved by increasing the number of controllable parameters in the scheme. We therefore consider a three-mode generation scheme based on particle-number measurements, shown in Fig. \ref{fig:scheme_3mode}.
\noindent In the scheme, three modes $S_1$, $S_2$ and $S_3$ in squeezed vacuum states are entangled via a quantum transformation $\hat{U}_3$. As before, we assume the entangling operation to be Gaussian, without specifying its explicit form. The resulting three-mode Gaussian state has the following wave function:
\begin{align} \label{wav_func_3}
    \phi_3 \left(x_1,x_2,x_3\right)= \mathcal{C}   e^ {-\frac{1}{2} \left(a_1 x_1^2+a_2 x_2^2+a_3
   x_3^2+2 b_{12} x_2 x_1+2 b_{13} x_3 x_1+2 b_{23} x_2 x_3\right)},
\end{align}
where$a_1$, $a_2$, $a_3$, $b_{12}$, $b_{13}$, $b_{23}$ are state parameters determined by the parameters of the generation scheme. The conditions on these parameters, as well as the explicit form of the normalization factor $\mathcal{C}$, are presented in Appendix \ref{append_G3}. Eq. (\ref{wav_func_3}) shows that the state is characterized by six independent parameters, which can be adjusted to generate approximate SSCSs.

The first and second modes are then measured using RNRDs. Conditioned on the measurement outcomes $n$ and $m$, the remaining mode is prepared in the output state whose wave function has the following form:
\begin{align} \label{out_3_Rz}
 |\Psi^{out,3}_{n,m} \left(r,\beta_1,\beta_2,\beta_3\right) \rangle= \hat{\mathcal{S}}(r) \left( \frac{1}{\sqrt{\mathcal{N}}}\sum _{k=0}^{\lfloor \frac{n+m}{2}\rfloor} \tilde{c}_k \left(\beta_1,\beta_2,\beta_3\right) |n+m-2k\rangle\right),
\end{align}
where $\mathcal{N}$ is the normalization of the state, and the expansion coefficients are given as follows:
\begin{align}
  \tilde{c}_k \left(\beta_1,\beta_2,\beta_3\right)=  \left(\frac{\beta_1}{2}\right)^k \sum_{l=0}^{\lfloor \frac{n+m}{2} \rfloor} \frac{(m+n-2 l)! \left(\frac{\beta_2}{\beta_1}\right)^l \, _2\tilde{F}_1\left(-2 l,-n;m+1-2 l;\beta_3\right)}{l! (k-l)! \sqrt{(m+n-2k)!}}.
\end{align}
Here $_2\tilde{F}_1\left(a,b;c ;x\right)$ is the regularized hypergeometric function, $r$ is the squeezing parameter, and $\beta_1$, $\beta_2$, and $\beta_3$ are parameters on which the expansion coefficients depend. The explicit relation between these parameters and those of the three-mode Gaussian state is provided in Appendix \ref{append_param_NG3}.

The Eq. (\ref{out_3_Rz}) shows that the generated state is a finite superposition of squeezed Fock states with definite parity. The parity of the generated state is determined by the sum of the detected particle numbers, $n+m$. Unlike the two-mode scheme, the expansion coefficients now depend on three parameters, $\beta_1$, $\beta_2$, and $\beta_3$. In the limiting cases $b_{13}=b_{12}=n=0$ and $b_{23}=b_{12}=m=0$, the state (\ref{out_3_Rz}) reduces to the state (\ref{out_Rz}), with the parameter $\beta _1\rightarrow z$. Thus, in the three-mode scheme, one can generate a richer class of states, enabling higher-fidelity approximations of SSCSs. We discuss this point in more detail below.

The probability of generating states in the scheme is given by the following expression:
\begin{align}
   \mathcal{P}_3(n,m)=\frac{4\pi ^{3/2} \mathcal{C}^2 \left| \alpha _2\right| {}^{2 m} \left| \alpha _1\right| {}^{2 n}}{ \sqrt{\Re(R)} |\alpha_5| ^{1+2 m+2 n}}\frac{m!}{n!} \sum \limits_{k=0}^{\lfloor \frac{n+m}{2}\rfloor} \left|\tilde{c}_k \left(\beta_1,\beta_2,\beta_3\right)\right|^2.
\end{align}
From the presented expression, one can conclude that the probability of generating a state in the three-mode case depends on all parameters of the scheme. To generate a given state, four parameters $r$, $\beta_1$, $\beta_2$, and $\beta_3$ are fixed, while the remaining two parameters can be tuned to maximize the generation probability.

\subsection{Fidelity of the generated SSCSs}
\subsubsection{General expression for the fidelity}
Having derived the explicit form of the generated state, we can now evaluate the fidelity of SSCS generation in the three-mode scheme. Since we are interested in SSCSs with real parameters $\alpha, R \in \mathbb{R}$, we restrict all parameters of the three-mode Gaussian state to be real. In this case, the fidelity between the generated state and an SSCS with amplitude $\alpha$ and squeezing parameter $R$ is given by the following expression:
\begin{multline} \label{fid_N3}
  F\left(|\Psi^{out,3}_{n,m} \left(r,\beta_1,\beta_2,\beta_3\right)\rangle,|\text{Scat}_{\pm}(\alpha,R)\rangle\right)= \frac{2 e^{\alpha^2 y}\sqrt{1-y^2}}{\mathcal{N}\left( e^{\alpha^2}+(-1)^{m+n}e^{-\alpha^2} \right)} \left(\frac{y-\beta_1}{2}\right)^{m+n}\\
  \times  \left(\sum_{p=0}^{\lfloor \frac{n+m}{2}\rfloor}\frac{\, _2\tilde{F}_1(-n,-2 p;m-2 p+1;\beta_3) \left(\frac{\beta_2}{y-\beta_1}\right)^p H_{m+n-2 p}\left(\frac{\alpha \sqrt{1-y^2}
   }{ \sqrt{2\left(y-\beta_1\right)}}\right)}{p!}\right)^2.
\end{multline}
Here, as before, we introduce the parameter $y=\tanh (R-r)$. By optimizing the parameters $\beta_1$, $\beta_2$, $\beta_3$, and the relative squeezing parameter $y$, we define the maximal fidelity as
\begin{align}
    \mathcal{F}_3\equiv \max_{y, \beta_1, \beta_2, \beta_3} \left[F\left(|\Psi^{out,3}_{n,m} \left(r,\beta_1,\beta_2,\beta_3\right)\rangle,|\text{Scat}_{\pm}(\alpha,R)\rangle\right)\right].
\end{align}
Unlike the two-mode case, the fidelity (\ref{fid_N3}) depends not only on the total detected particle number $n+m$, but also on the individual values of $n$ and $m$ (through the hypergeometric function). Consequently, the optimization can also be performed over the detected particle numbers $n$ and $m$. In physical terms, this means that to generate an SSCS with the highest possible fidelity, one must choose a generation scheme and wait for specific detection outcomes $n$ and $m$.

\subsubsection{The case of detecting zero particles at one of the detectors} 
Before presenting a general analysis of the SSCS generation fidelity in the scheme under consideration, let us first examine several special cases. For ease of comparison, we fix the total detected particle number $n+m=N$ and examine different distributions of detected particles between the two PNRDs. We begin with the case in which one of the detectors registers zero particles. For $n=0$ and $m=N$, the fidelity is given by the following expression:
\begin{align} \label{fid_n0}
\mathcal{F}_3 \Big|_{n=0}=  \max_{y,\beta_1, \beta_2} \left[\frac{2 e^{\alpha^2 y} \sqrt{1-y^2}  \left(\frac{y-\beta_1-\beta_2}{2}\right)^N H_m^2\left(\frac{\alpha}{\sqrt{2}}\frac{\sqrt{1-y^2}}{\sqrt{y-\beta_1-\beta_2}}\right)}{N! \left((-1)^N e^{-\alpha^2}+e^{\alpha^2}\right) \,
   _2F_1\left(\frac{1-N}{2},-\frac{N}{2};1;(\beta_1+\beta_2)^2\right)}\right].
\end{align}
It is important to note that the fidelity in the symmetric case $m=0$ and $n=N$ coincides with Eq. (\ref{fid_n0}) up to the change of variables: $\beta_2 \rightarrow \beta_2 \beta_3^2$. This follows from the symmetry of the problem. Since we consider a general Gaussian state rather than a specific implementation of the generation scheme, the state parameters can always be redefined such that the fidelity remains invariant under the exchange $n\leftrightarrow m$. Therefore, in what follows, we restrict our analysis to fixed values of $n$, with $m=N-n$.

Eq. (\ref{fid_n0}) shows that the fidelity is identical to that of the two-mode case (\ref{fid}) up to the change $z \rightarrow -\beta_1-\beta_2$. Consequently, the maximal fidelity in this case is identical to that obtained in the two-mode scheme.  Furthermore, Eq. (\ref{fid_n0}) shows that maximizing the fidelity requires fixing y and the sum  $\beta_1+\beta_2$. This means that four free parameters remain in the scheme, which may be used either to simplify the experimental implementation or to maximize the generation probability of a given SSCS.

\subsubsection{The case of detecting one particle at one of the detectors}
Let us now consider the case of detecting one particle at one of the detectors. In the case $n=1$ and $m=N-1$, the fidelity of SSCS generation is given by the following expression:
\begin{multline}
\mathcal{F}_3\Big|_{n=1}= \max _{y,\beta_1,\beta_2,\beta_3}\left[\frac{2e^{\alpha^2 y}\sqrt{1-y^2} \left(\frac{y-\beta _1-\beta _2}{2}\right)^{N}}{N! \left(e^{\alpha^2}+(-1)^{N}
   e^{-\alpha^2}\right) } \left(\frac{1}{\sqrt{\mathcal{N}_1}}H_{N}\left(\frac{\alpha \sqrt{1-y^2}}{\sqrt{2} \sqrt{y-\beta _1-\beta _2}}\right) \right. \right. \\
 \left. \left. +\frac{2 \beta _2
   \left(\beta _3-1\right) (N-1) }{\sqrt{\mathcal{N}_1}\left(y-\beta _1-\beta _2\right)}H_{N-2}\left(\frac{\alpha \sqrt{1-y^2}}{\sqrt{2} \sqrt{y-\beta _1-\beta _2}}\right)\right)^2\right],
\end{multline}
where $\mathcal{N}_1$ is the normalization factor, whose explicit expression is presented in Appendix \ref{append_Norm}. It follows from the above expression that one of the conditions under which the maximal fidelity is attained is $\beta _2 \left(\beta _3-1\right)=0$. In this case, the maximal fidelity coincides with that in the two-mode case. Although this is not the only parameter choice leading to the maximal fidelity, the maximal fidelity never exceeds that in the two-mode case.

Thus, when one particle is detected by one of the PNRDs, the maximal fidelity of SSCS generation coincides with that in the two-mode case. Therefore, no fidelity gain is achieved compared with the two-mode scheme. Nevertheless, as in the case where one detector registers zero particles, several free parameters remain in the scheme.These can be used either to maximize the generation probability or to simplify the experimental implementation.

\subsubsection{The cases of detecting two and a greater number of particles at one of the detectors}
We now consider the case where one detector registers two particles ($n=2$), while the other registers $m=N-2$.The fidelity in this case is given by the following expression:
\begin{multline}
    \mathcal{F}_3\Big|_{n=2}=\max_{y,\beta_1, \beta_2, \beta_3} \left[\frac{2 e^{\alpha^2 y} \sqrt{1-y^2} \left(\frac{y-\beta _1-\beta _2}{2}\right)^{N}}{N! \left(e^{\alpha^2}+(-1)^{N} e^{-\alpha^2}\right)}\left(\frac{H_{N}\left(\frac{\alpha \sqrt{1-y^2}}{\sqrt{2} \sqrt{y-\beta _1-\beta _2}}\right)}{\sqrt{\mathcal{N}_2}} \right. \right. \\
    \left. \left. +\frac{2 \beta _2 \left(\beta _3-1\right) \left(\beta _3+2N-3\right) H_{N-2}\left(\frac{\alpha \sqrt{1-y^2}}{\sqrt{2} \sqrt{y-\beta _1-\beta
   _2}}\right)}{\sqrt{\mathcal{N}_2} \left(y-\beta _1-\beta _2\right)} +\frac{4 \beta _2^2 \left(\beta _3-1\right)^2 (N-3) (N-2) H_{N-4}\left(\frac{\alpha \sqrt{1-y^2}}{\sqrt{2} \sqrt{y-\beta _1-\beta _2}}\right)}{\sqrt{\mathcal{N}_2} \left(y-\beta _1-\beta
   _2\right){}^2}\right)^2\right],
\end{multline}
where $\mathcal{N}_2$ is the normalization factor, whose explicit form is presented in Appendix \ref{append_Norm}. The above expression shows that the fidelity optimization requires independent control of the three coefficients  in front of the Hermite polynomials. This is possible because the three parameters $\beta_1$, $\beta_2$, and $\beta_3$ uniquely determine these three coefficients.

Increasing the detected particle number $n$ at one of the detectors introduces additional Hermite polynomials into the fidelity expression. For example, when $n=3$, the expression for the fidelity already contains four Hermite polynomials. The corresponding expression is given in Appendix \ref{append_fid_N3_3}. Since the number of controllable parameters remains equal to three, cases with $n \geqslant 3$ cannot achieve higher fidelities than the case $n=2$ for the same total detected particle number $N$. Therefore, the maximal fidelity is achieved when two particles are detected by one of the PNRDs. In all other cases, the maximally achievable fidelity of the SSCS will be no greater.

\begin{figure}[H]
    \centering
    \includegraphics[width=0.6\linewidth]{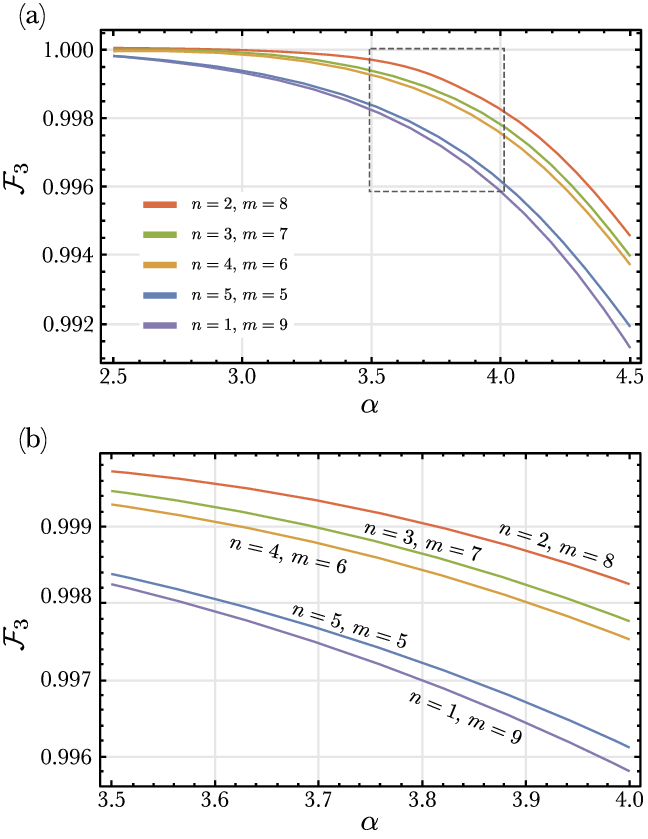}
    \caption{(a) Fidelity of the generated SSCSs as a function of their amplitude $\alpha$. In the figure, different colors denote different cases of detecting the number of particles $n+m=10$. (b) Enlarged image of the region outlined by the dashed line in Fig. \ref{fig:tm_10_sum} (a).}
    \label{fig:tm_10_sum}
\end{figure}

As an illustration, Fig. \ref{fig:tm_10_sum} shows the fidelity of the SSCSs generated in the three-mode scheme as a function of their amplitude $\alpha$ for different cases of detecting a total number of particles $N=10$.
\noindent The figure shows that the maximal fidelity is achieved for the case of detecting two particles at one of the detectors. The lowest fidelity among the cases shown coincides with that in the two-mode case. Such fidelity is achieved when detecting one particle or zero particles at one of the detectors. Although the figure is shown for $N=10$, the same qualitative behavior is observed for other values of the total detected particle number.

Thus, detecting two particles at one of the PNRDs yields the largest fidelity gain of the three-mode scheme over the two-mode one.

\subsubsection{The generation parameter $\beta_3=1$}
Let us consider another important special case. Instead of fixing the detected particle numbers, we fix the parameter of the generated state to $\beta_3=1$. In this case, the SSCS generation fidelity is given by the following expression:
\begin{align}
\mathcal{F}\Big|_{\beta_3=1} =  \frac{2 e^{\alpha^2 y} \sqrt{1-y^2}  \left(\frac{y-\beta_1-\beta_2}{2}\right)^{N} H_{N}^2\left(\frac{\alpha}{\sqrt{2}}\frac{\sqrt{1-y^2}}{\sqrt{y-\beta_1-\beta_2}}\right)}{N! \left(e^{\alpha^2}+(-1)^{N} e^{-\alpha^2}\right) \,
   _2F_1\left(\frac{1-N}{2},-\frac{N}{2};1;(\beta_1+\beta_2)^2\right)}.
\end{align}
We see that, in this case, the fidelity expression coincides with Eq. (\ref{fid_n0}) up to the substitution $n\rightarrow n+m$ and $z\rightarrow -\beta_1-\beta_2$. Consequently, the maximal achievable fidelity is identical to that in the two-mode case.

\begin{figure}[H]
    \centering
    \includegraphics[width=0.45\linewidth]{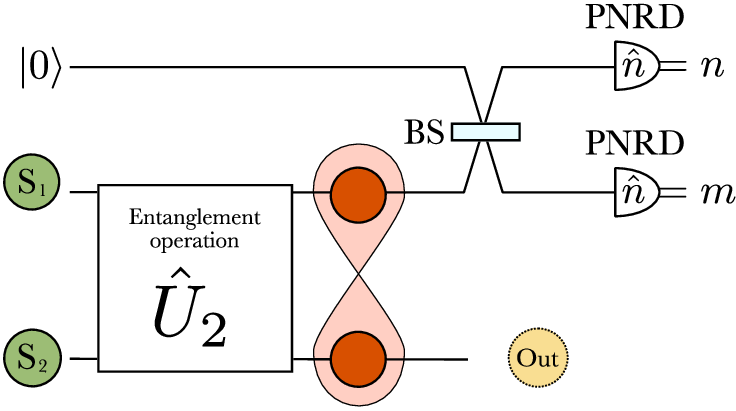}
    \caption{A special case of the three-mode generation scheme, in which one of the modes is in the vacuum state. In the figure,  $\text{S}_1$  and $\text{S}_2$ are two squeezed vacuum states, $\hat{U}_2$ is some two-mode entangling Gaussian operation, BS is a beam splitter, PNRD is a photon-number-resolving detector, and Out is the output state}
    \label{fig:2mode+vac}
\end{figure}

Having analyzed this generation case, one can conclude that the parameter $\beta_3=1$ is achieved in the schemes shown in Fig. \ref{fig:2mode+vac}.
\noindent The presented three-mode generation scheme reduces to a two-mode one, in which the measured mode is split by a beam splitter into two channels, each measured by a PNRD. The third input mode is prepared in the vacuum state.

The found relation between the three-mode and two-mode schemes allows us to compare the generation probabilities in these two cases. Since for $\beta_3=1$ the generation fidelity depends only on the total detected particle number $m+n=N$, and not on the specific values of $m$ and $n$, the probability of generating an SSCS with a given fidelity is the sum of probabilities:
\begin{align}
    \sum_{\substack{n,m \\ n+m=N}} \mathcal{P}_3(n,m)\Big|_{\beta_3=1}=\mathcal{P}_2(N),
\end{align}
which is precisely the probability of generating the same SSCS in the two-mode scheme upon detecting $N$ particles. Moreover, for any $n>1$ and $m>1$ that together give $n+m=N$, the following estimate holds:
\begin{align}
\mathcal{P}_3(n>1,m>1)\Big|_{n+m=N} \leqslant \mathcal{P}_2(N).
\end{align}
This inequality means that in the three-mode scheme, the probability of generating an SSCS upon detecting certain particle numbers $n>1$ and $m>1$ ($n+m=N$) at the two detectors never exceeds the generation probability in the two-mode case upon detecting $N$ particles. This result has a simple physical interpretation. For example, detecting a total of ten particles across two detectors is more likely than detecting exactly two particles in one detector and eight in the other.

\subsection{Growth of the amplitude of SSCSs generated in the three-mode scheme}
\begin{figure}[H]
    \centering
    \includegraphics[width=0.5\linewidth]{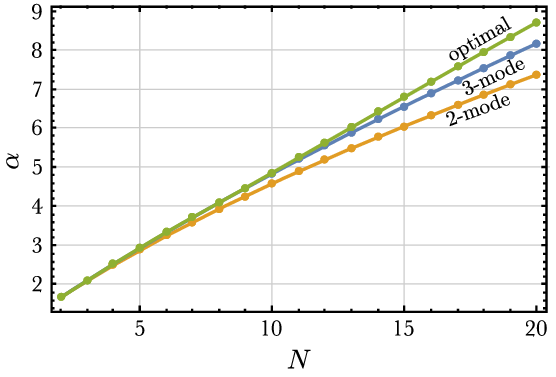}
    \caption{Dependence of the maximal SSCS amplitude on the detected number of particles. In the figure, different colors denote three cases: the blue curve corresponds to generation in the three-mode scheme, the yellow curve to the two-mode scheme, and the green curve to the maximal amplitude obtained in the optimal case.}
    \label{fig:fid_comp}
\end{figure}

Having examined these special cases, we now analyze how the generated SSCS amplitude increases with the detected particle number in the three-mode generation scheme. Fig. \ref{fig:fid_comp} presents plots of the dependence of the maximal SSCS amplitude , generated with a fidelity of $0.99$,  on the detected number of particles.
\noindent The figure shows that for a small detected number of particles, the difference between the various cases is minimal. As the detected number of particles grows, the scheme with a larger number of modes is better suited for generating SSCSs, since it contains a larger number of controllable parameters. In particular, up to $N=10$ the SSCS amplitude in the scheme with three modes practically coincides with the maximal achievable amplitude. As the detected particle number increases, the achievable amplitude in the three-mode scheme increases more rapidly than in the two-mode scheme. For example, at $N=20$, an SSCS generated with fidelity $0.99$ has an amplitude one unit greater than in the scheme with two modes.

It is important to note, however, that the maximal fidelity (and amplitude) in the three-mode scheme is achieved when two particles are detected by one PNRD and $N-2$ by the other. As discussed above, in this case the probability of generating an SSCS will be smaller than the generation probability in the two-mode case.

Thus, increasing the number of modes in the generation scheme improves the achievable fidelity (and amplitude) by providing additional controllable parameters. At the same time, it reduces the generation probability, since the specific measurement outcome at each detector must be obtained.

\section{Conclusion}
In this work, we considered the generation of SSCSs in schemes based on the measurement of Gaussian states by means of PNRDs. Based on a general description of the generated states, we determined the maximal fidelity with SSCSs. It was shown that, in the optimal case, the maximal achievable fidelity increases with the detected particle number, but so does the number of parameters that must be optimized to attain it. Since the number of controllable parameters is determined by the number of modes, high-fidelity generation of large-amplitude SSCSs requires multimode generation schemes.

We then analyzed the general two-mode generation scheme. We found that even in the two-mode scheme, for a small detected number of particles, one can generate SSCSs with fidelities close to the optimal values. For example, upon detecting seven particles, the relative improvement of the optimal case over the two-mode scheme is less than one percent. However, this improvement increases with the detected particle number, and therefore, for generating SSCSs with large amplitudes, the two-mode scheme will be less efficient. This is related to the small number of controllable parameters in the two-mode scheme.

To improve the characteristics of the generated SSCSs, we considered a three-mode generation scheme. The analysis revealed that such a scheme has a larger number of controllable parameters, which leads to the generation of SSCSs with large amplitudes. It was demonstrated that, upon detecting one particle or zero particles at one of the detectors in such a scheme, the fidelity of the generated SSCSs does not exceed the maximal fidelity in the two-mode case. In addition, we showed that the highest fidelity is achieved when two particles are detected by one of the detectors.

We further found that, when more than one particle is detected by each PNRD, the three-mode scheme has a lower generation probability than the corresponding two-mode scheme. Consequently, the fidelity improvement provided by the three-mode scheme comes at the expense of a reduced success probability.

In conclusion, increasing the number of modes in the generation scheme leads to an increase in the amplitude of the generated SSCSs with a given fidelity, and to a decrease in the generation probability. Therefore, in real experimental setups, when generating SSCSs, one must seek a certain compromise between these two parameters.
\vspace{0.5 cm}

This research was supported by the Theoretical Physics and Mathematics Advancement Foundation "BASIS" (Grant No. 24-1-3-14-1).

\bibliography{bibliography}
%%%%%%%%%%%%%%%%%%%%%%%%%%%%%%%%%%%%%%%%%%%%%%%%%%%%%%%%%%%%%%%

\appendix
\section{Parameters of the three-mode Gaussian state} \label{append_G3}
Normalization of the three-mode Gaussian state:
\begin{align}
    \mathcal{C}=\frac{\sqrt[4]{\Re\left[a_1\right] \Re\left[a_2\right] \Re\left[a_3\right]+2 \Re\left[b_{13}\right] \Re\left[b_{23}\right]
   \Re\left[b_{12}\right]-\Re\left[a_3\right] \Re\left[b_{12}\right]^2-\Re\left[a_2\right] \Re\left[b_{13}\right]^2-\Re\left[a_1\right]
   \Re\left[b_{23}\right]^2}}{\pi ^{3/4}}.x
\end{align}
The parameters of the three-mode state satisfy the following conditions: $\Re [a_1]>0$, $\Re[a_2]>0$, $\Re[a_3]>0$, $\Re\left[a_1\right] 
\Re\left[a_2\right]-\Re\left[b_{12}\right]^2>0$, $\Re\left[a_2\right] \Re\left[a_3\right]-\Re\left[b_{23}\right]^2>0$, $\Re\left[a_1\right] \Re\left[a_3\right]-\Re\left[b_{13}\right]^2>0$ and $\Re\left[a_1\right] \Re\left[a_2\right]
   \Re\left[a_3\right]+2 \Re\left[b_{13}\right] \Re\left[b_{23}\right]
   \Re\left[b_{12}\right]-\Re\left[a_3\right] \Re\left[b_{12}\right]{}^2-\Re\left[a_2\right]
   \Re\left[b_{13}\right]^2-\Re\left[a_1\right]
   \Re\left[b_{23}\right]^2>0$.

\section{Parameters of the state generated in the three-mode scheme} \label{append_param_NG3}
The relation of the parameters of the state generated in the three-mode scheme based on particle-number measurements to the parameters of the three-mode Gaussian state has the following form:
 \begin{align}
   &\beta_1=1-\frac{\alpha_5 \left(\left(\alpha _2+\alpha _3\right) \alpha _3 \alpha_5+ \left(\alpha _1+\alpha _4\right)\alpha _4
   \alpha _5-4 \alpha _2 \alpha _3-4 \alpha _1 \alpha _4\right)}{\left(\alpha _2 \alpha _3+\alpha _1 \alpha _4\right) \left(\alpha _2 \left(\alpha _2+\alpha _3\right)+\alpha _1 \left(\alpha _1+\alpha
   _4\right)\right)} \Re[\rho],\\
   &\beta_2=\frac{\alpha _4^2 \alpha_5^2 \Re[\rho]}{\alpha _2^2 \left(\alpha _2 \alpha _3+\alpha _1 \alpha _4\right)},\\
   &\beta_3=-\frac{\alpha _2 \alpha _3}{\alpha _1 \alpha _4},
\end{align}
where the following abbreviations are introduced:
\begin{align}
    &\alpha _1=\left(a_2+1\right) b_{13}-b_{12} b_{23},\\
   &\alpha _2=\left(a_1+1\right) b_{23}-b_{12} b_{13},\\
   &\alpha _3=(1-a_1) b_{23}+b_{12} b_{13},\\
   &\alpha _4=(1-a_2) b_{13}+b_{12} b_{23},\\
  & \alpha_5=\left(a_1+1\right) \left(a_2+1\right)-b_{12}^2,\\
 &\rho=a_3-\frac{\left(a_2+1\right) b_{13}^2+\left(a_1+1\right) 
 b_{23}^2-2 b_{12} b_{23} b_{13}}{\left(a_1+1\right) \left(a_2+1\right)-b_{12}^2}.
 \end{align}
The complex squeezing parameter $r=|r|e^{i\arg [r]}$ of the generated state is defined as follows:
\begin{align}
    \cosh (|r|)=\frac{\left|1+\rho\right| }{2 \sqrt{\Re\left[\rho\right]}},\quad \tan (\arg [r])=\frac{2 \Im\left[\rho\right]}{\left|\rho\right|^2-1}. 
\end{align}

\section{Normalization factors for the SSCS generation fidelity in various special cases} \label{append_Norm}
The normalization factor in the case of detecting one particle at one of the detectors:
\begin{multline}
  \mathcal{N}_1= \, _2F_1\left(\frac{1-N}{2},-\frac{N}{2} ;1;\left(\beta _1+\beta _2\right){}^2\right)\\
  +(N-1)\beta _2 \left(\beta _3-1\right) \left(\beta
   _1+\beta _2\right) \, _2F_1\left(\frac{2-N}{2},1-\frac{N-1}{2};2;\left(\beta _1+\beta _2\right){}^2\right)\\
   +\frac{\beta _2^2 \left(\beta _3-1\right)^2 (N-1)}{N} \,
   _2F_1\left(\frac{2-N}{2},1-\frac{N-1}{2};1;\left(\beta _1+\beta _2\right){}^2\right).
\end{multline}

The normalization factor for the fidelity in the case of detecting two particles at one of the detectors:
\begin{multline}
    \mathcal{N}_2=\,_2F_1\left(\frac{1-N}{2},-\frac{N}{2};1;\left(\beta _1+\beta _2\right){}^2\right)\\
    +\beta _2 \left(\beta _3-1\right) \left(\beta _1 \left(\beta _3+2 N-3\right)-\beta _2 \left(\beta _3-2 N+1\right)\right) \, _2F_1\left(\frac{3-N}{2},1-\frac{N}{2};2;\left(\beta _1+\beta _2\right){}^2\right)\\
   +\frac{\beta _2^3 \left(\beta _3-1\right){}^3 (N-3) (N-2) \left(2 \beta _2 \left(\beta _3+N-2\right)+\beta _1 \left(\beta _3+2
   N-3\right)\right)\, _2F_1\left(\frac{5-N}{2},2-\frac{N}{2};2;\left(\beta _1+\beta _2\right){}^2\right)}{N(N-1)} \\
   +\frac{\beta _2^2 \left(\beta _3-1\right){}^2 \left(\beta _3 \left(\beta _3+4 N-6\right)+2 N (3 N-7)+9\right) \,
   _2F_1\left(\frac{3-N}{2},1-\frac{N}{2};1;\left(\beta _1+\beta _2\right){}^2\right)}{N(N-1)}\\
   +\frac{\beta _2^4 \left(\beta _1+\beta _2\right){}^2 \left(\beta _3-1\right){}^4 (N-5) (N-4) (N-3) (N-2) \,
   _2F_1\left(\frac{7-N}{2},3-\frac{N}{2};3;\left(\beta _1+\beta _2\right){}^2\right)}{8 (N-1) N}.
\end{multline}

\subsection{Fidelity of SSCS generation in the three-mode case upon detecting three particles at one of the detectors} \label{append_fid_N3_3}
The expression for the fidelity when three particles are detected at one of the detectors:
\begin{multline}
 \mathcal{F}\Big|_{n=3}=\max_{y,\beta_1,\beta_2,\beta_3} \left[\frac{2 e^{\alpha^2 y} \sqrt{1-y^2} \left(\frac{y-\beta _1-\beta _2}{2}\right)^{N}}{N! \left(e^{\alpha^2}+(-1)^N e^{-\alpha^2}\right)}\left(\frac{1}{\sqrt{\mathcal{N}_3}}H_{N}\left(\frac{\alpha \sqrt{1-y^2}}{\sqrt{2} \sqrt{y-\beta _1-\beta _2}}\right) \right. \right.\\
 +\left. \frac{6 \beta _2 \left(\beta _3-1\right) \left(\beta _3+N-2\right) H_{N-2}\left(\frac{\alpha \sqrt{1-y^2}}{\sqrt{2} \sqrt{y-\beta
   _1-\beta _2}}\right)}{\sqrt{\mathcal{N}_3}(y-\beta _1-\beta _2)}+\frac{12 \beta _2^2 \left(\beta _3-1\right)^2 (N-3) \left(\beta _3+N-3\right) H_{N-4}\left(\frac{\alpha \sqrt{1-y^2}}{\sqrt{2} \sqrt{y-\beta _1-\beta
   _2}}\right)}{\sqrt{\mathcal{N}_3}\left(y-\beta _1-\beta _2\right){}^2}\right. \\
   \left. \left. + \frac{8 \beta _2^3 \left(\beta _3-1\right)^3 (N-5) (N-4) (N-3) H_{N-6}\left(\frac{\alpha \sqrt{1-y^2}}{\sqrt{2} \sqrt{y-\beta _1-\beta _2}}\right)}{\sqrt{\mathcal{N}_3}\left(y-\beta _1-\beta
   _2\right){}^3} \right)^2 \right],
\end{multline}
where $\mathcal{N}_3$ is the normalization factor.
\end{document}